\documentclass[preprint,authoryear,12pt]{elsarticle}

\usepackage[utf8]{inputenc}
\usepackage{amsmath}
\usepackage{natbib}
\usepackage{graphicx}

\usepackage{color} 
\usepackage{lineno}
\usepackage{hyperref}
\hypersetup{
    bookmarks=true,         %
    unicode=false,          %
    pdftoolbar=true,        %
    pdfmenubar=true,        %
    pdffitwindow=false,     %
    pdfstartview={FitH},    %
    pdftitle={My title},    %
    pdfauthor={Author},     %
    pdfsubject={Subject},   %
    pdfcreator={Creator},   %
    pdfproducer={Producer}, %
    pdfkeywords={keywords}, %
    pdfnewwindow=true,      %
    colorlinks=true,       %
    linkcolor=blue,          %
    citecolor=blue,        %
    filecolor=magenta,      %
    urlcolor=cyan           %
}

\journal{Cognitive Systems Research}

\begin{document}

\begin{frontmatter}

\title{When to (or not to) trust intelligent machines: Insights from an evolutionary game theory analysis of trust in repeated games}

\author[1]{The Anh Han}
\ead{theanhhan.vn@gmail.com}
\author[2]{Cedric Perret}
\ead{cedric.perret.research@gmail.com}
\author[3]{Simon T. Powers\corref{cor1}}
\ead{S.Powers@napier.ac.uk}
\cortext[cor1]{Corresponding author}
\address[1]{Teesside University}
\address[2]{Teesside University}
\address[3]{Edinburgh Napier University}

\begin{abstract}
The actions of intelligent agents, such as chatbots, recommender systems, and virtual assistants are typically not fully transparent to the user. Consequently, using such an agent involves the user exposing themselves to the risk that the agent may act in a way opposed to the user's goals. It is often argued that people use trust as a cognitive shortcut to reduce the complexity of such interactions. Here we formalise this by using the methods of evolutionary game theory to study the viability of  trust-based strategies in repeated games. These are reciprocal strategies that cooperate as long as the other player is observed to be cooperating. Unlike classic reciprocal strategies, once mutual cooperation has been observed for a threshold number of rounds they stop checking their co-player's behaviour every round, and instead only check with some probability. By doing so, they reduce the \textit{opportunity cost} of verifying whether the action of their co-player was actually cooperative. We demonstrate that these trust-based strategies can outcompete strategies that are always conditional, such as Tit-for-Tat, when the opportunity cost is non-negligible. We argue that this cost is likely to be greater when the interaction is between people and intelligent agents, because of the reduced transparency of the agent. Consequently, we expect people to use trust-based strategies more frequently in interactions with intelligent agents. Our results provide new, important insights into the design of mechanisms for facilitating interactions between humans and intelligent agents, where trust is an essential factor.

\end{abstract}


\end{frontmatter}

\section{Introduction}
Artificial intelligence is undoubtedly becoming more integrated into our every day lives. While much attention has recently been paid to deep machine learning, intelligent agents that exhibit goal directed behaviour \citep{Wooldridge:2009:a} have also come of age. These range from purely software systems such as videogame characters or chatbots, through to cyberphysical systems such as smart fridges or autonomous vehicles. We are delegating more and more aspects of our daily lives to these agents, from the virtual sales agent that recommends products and services to us on an e-commerce website \citep{Beldad:2016:a}, to the intelligent virtual assistant (e.g. Amazon Alexa,  Apple Siri, Google Home) that plans our route to work and orders goods and services for us on command \citep{Chung:2017:a}. But in all of these cases, the operation of the agent is not fully transparent to the end user. Although research in explainable AI is beginning to address these issues \citep{Nunes:2017:a}, it seems unlikely that a user will ever be able to get complete information about how and why the agent has taken a particular decision. Consequently, using such an agent, and accepting its recommendations, necessarily involves the user placing some degree of trust in the agent. In the broadest sense, trust is willingness to take risk under uncertainty \citep{Luhmann:1979:a}. Here the risk is that the agent will act in a way opposed to our own goals, and the uncertainty comes from us lacking complete information about the behaviour of the agent to be able to ascertain this.

For example, consider again the virtual sales agent operating on the website of an e-commerce company \citep{Chattaraman:2012:a}, which sells products to customers based on the information that it learns about the customer through a chat dialogue, i.e. an agent-based recommender system \citep{Pu:2007:a,Yoo:2012:a,Jugovac:2017:a}. When a customer interacts with this virtual sales agent it does not have complete information about why product A from company X is being recommended as opposed to product B from company Y \citep{Grabner-Kraeuter:2002:a}. Thus, if the customer is going to use the virtual sales agent, they must take some degree of risk, for example, that the virtual sales agent recommends more expensive products, or those from manufacturers that the seller has a preferential relationship with, or does not provide full information about the quality of the product. Without a full understanding of the virtual sales agent's source code, the specifications of the alternative products, and the relationships between the sellers and manufacturers \citep{Akerlof:1970:a,Mahadevan:2000:a,Lewis:2011:a}, some degree of risk and hence trust must be involved \citep{Luhmann:1979:a,Grabner-Kraeuter:2002:a}. Similarly, when a virtual assistant gives us directions, we do not have complete information either about the route planning algorithm that it is using, or about relevant environmental conditions such as traffic levels. Again, this means that the use of such systems necessarily involves some degree of risk and hence trust.

This raises the question: how will people behave when interacting with these kinds of intelligent agents? How will they handle the complexity of the interaction? Ultimately, this question will need to be answered by empirical work. However, to guide the empirical work it is necessary to generate hypotheses about how we expect people to behave. Because intelligent agents exhibit goal directed behaviour, and their goals (as programmed by their designers) may be in conflict with the goals of their users, evolutionary game theory (EGT)  \citep{MaynardSmith:1982:a,key:Sigmund_selfishnes} provides a suitable formal framework for modelling the strategic interaction and understanding behavioural dynamics \citep{Shoham:2008:a}. This is because not only is the interaction strategic, but there is empirical evidence that people use a standard set of social scripts whether they are interacting with a person or a machine in a particular social situation \citep{Nass2000MachinesComputers}. This suggests that predictions from game theoretical studies about human behaviour in traditional (e-)commerce, for example (e.g. \citealt{Laaksonen:2009:a,Dahlstrom:2014:a}), can also be useful when the interaction is between a human and an intelligent agent representing another entity (individual, firm, organisation), rather than with that entity directly.   

In light of this, we propose that the types of interaction discussed above can be modelled as repeated games between the user and the agent (acting to fulfil the goals of its designer). Moreover, in important cases the actions available to the agent and the user correspond to ``cooperate'' and ``defect''. Cooperation between players represents both the user and agent behaving honestly, reliably and transparently with each other. For example, cooperation would be a virtual sales agent selling products that match the preferences that the user has revealed in the conversation, while defection might correspond to trying to upsell products or warranties. On the side of the user, cooperation could represent continued use of the agent, which benefits the seller by reducing their opportunity costs of answering customer enquiries themselves. Defection would then represent refusing to use the agent and instead speaking directly to a human sales advisor. 

The folk theorem of repeated games tells us that the key to cooperative outcomes, which benefit both sides, is sufficient information for the players to be able to condition their actions on the past actions of the other player(s) \citep{Fudenberg:1986:a}. This allows for reciprocal strategies, e.g. cooperate if the other player cooperated in the previous interaction, as exemplified by the Tit-for-Tat strategy \citep{key:axelrod84}. However, the use of reciprocal strategies necessarily carries an opportunity cost. Part of this comes from devoting cognitive resources to remembering a history of past actions, and processing this when deciding how to act. But in addition to this, reciprocal strategies also involve \textit{verifying} whether the observed action of the other player actually was cooperative or not. In traditional face-to-face interactions between humans verifying whether the other player cooperated might involve, for example, checking the quality and specification of goods that have been purchased, or that the correct amount of change has been given. However, these costs are usually assumed to  be  low compared to the benefit and cost of cooperation \citep{ho1996finite,Imhof2005EvolutionaryDefection,han2013intention}, and are mostly omitted in (evolutionary) game theoretic models  \citep{mcnally2012cooperation,martinez2015apology,garcia2018no,hilbe2017memory}. But the move to interactions over the internet increases these costs \citep{Grabner-Kraeuter:2002:a}, since the increased separation in space and time over the course of the interaction makes verifying the action of the other player more costly. The move to interacting with intelligent agents increases these costs even more, since the interaction becomes less transparent to the user, and artificial agents have limited capacity to explain their action compared to humans. This issue is increasingly more  relevant when considering hybrid societies of humans and intelligent agents \citep{paiva2018engineering,santos2019evolution}.

It is often argued that humans use trust as a cognitive shortcut, to reduce the complexity of the interaction that they need to reason about \citep{Luhmann:1979:a,Grabner-Kraeuter:2002:a,Petruzzi:2014:a}. In this paper, we formalise this in EGT by introducing trust-based strategies in repeated games, and study their evolutionary viability when competing with other strategies in repeated games, in a similar fashion to \cite{Imhof2005EvolutionaryDefection}. Unlike traditional  Tit-for-Tat, trust-based strategies  only check a co-player's actions occasionally after a trust threshold has been reached, i.e. after their co-player has cooperated for a certain number of rounds. By doing so, they reduce the opportunity cost of verifying the action of their co-player every round. We demonstrate that trust-based strategies can be more successful than Tit-for-Tat when the opportunity cost of using a conditional strategy is non-negligible. Moreover, one may ask under what kinds of interaction or business at hand are trust-based strategies more likely to be used by the parties involved? For instance, will users trust in a chatbot to handle highly important interactions such as a multi-million dollars transition? 
We show that trust-based strategies are most successful when the interaction is of intermediate importance, and the interaction is repeated over many rounds. These results provide game theoretic support for the theory that humans use trust to reduce the complexity of interactions, and suggest that people are likely to behave even more in this manner when interactions are with intelligent agents, since the opportunity costs of verifying the actions of intelligent agents are likely to be greater.

\section{Models and Methods}
\subsection{Models}
We consider a population of constant size $N$. At each time step or generation, a random pair of players are chosen to play with each other. 
\subsection{Interaction between Individuals} 
Interactions are modelled as a symmetric two-player Prisoner's Dilemma game, defined by the following  payoff
matrix (for row player) 
\[
 \bordermatrix{~ & C & D\cr
                  C & R & S \cr
                  D & T & P  \cr
                 }.
\]
A player who chooses to cooperate (C) with someone who defects (D) receives the sucker's payoff $S$, whereas the defecting player gains the temptation to defect, $T$. Mutual cooperation (resp., defection) yields the reward $R$ (resp., punishment P) for both players.  
Depending on the ordering of these four payoffs, different social dilemmas arise \citep{key:macy2002, key:santos2006}. Namely, in this work we are concerned with the Prisoner's Dilemma (PD), where $T > R > P > S$. 
In a single round, it is always best to defect, but cooperation may be rewarded if the game is repeated. In repeated PD, it is also required that mutual cooperation is preferred over an equal probability of unilateral cooperation and defection ($2R > T +S$); otherwise alternating between cooperation and defection would lead to a higher payoff than mutual cooperation. For convenience and a clear representation of results, we later mostly use the  Donation game \citep{key:Sigmund_selfishnes}---a famous  special case of the PD---where $T = b, \ R = b-c, \ P = 0, \ S = -c$, satisfying that $b > c > 0$,  where b and c stand respectively for ``benefit" and ``cost" (of cooperation). 

In addition, in order to understand how the duration of the interaction or business at hand impacts the evolutionary viability of trust-based strategies in relation to others, we model how  important or beneficial an interaction is using   parameter $\gamma > 0$ \citep{han2013good}. Hence, the payoff matrix becomes  
\[
 \bordermatrix{~ & C & D\cr
                  C & \gamma R & \gamma S \cr
                  D &  \gamma T &  \gamma P  \cr
                 }.
\]

In a population of $N$ individuals interacting via a repeated (or iterated) Prisoner's dilemma, whenever two specific strategies are present in the population, say \textbf{A} and \textbf{B}, the fitness of an individual with a strategy \textbf{A}  in a population with $k$  \textbf{A}s and $(N-k)$ \textbf{B}s can be written as  
\begin{equation} 
\label{eq:PayoffA} 
\Pi_A(k) =\frac{1}{r(N-1)}\sum_{j=1}^{r}[(k-1)\pi_{A,A}(j) + (N-k)\pi_{A,B}(j)],
\end{equation} 
where  $\pi_{A,A}(j)$ ($\pi_{A,B}(j)$) stands for the payoff obtained from a round $j$ as a result of their mutual behavior of an \textbf{A} strategist in an interaction with a \textbf{A} (\textbf{B}) strategist (as specified by the payoff matrix above), and $r$ is the total number of rounds of the Prisoner's dilemma. As usual, instead of considering a fixed number of rounds, upon completion of each round, there is a probability $w$ that yet another round of the game will take place, resulting in an average number of  $r =(1-w)^{-1}$ rounds per interaction \citep{key:Sigmund_selfishnes}. In the following, all values of $\Pi$ will be computed analytically. %

\subsection{Strategies in IPD and the opportunity cost}
The repeated (or iterated) PD is usually known as a story of tit-for-tat (TFT), which won both Axelrod's tournaments  \citep{key:axelrod84,key:axelrod81}. \textit{TFT} starts by cooperating, and does whatever the opponent did in the previous round. It will cooperate if the opponent cooperated, and will defect if the opponent defected. 

As a conditional strategy, TFT incurs an additional opportunity cost, denoted by $\epsilon$, compared to the unconditional strategies, namely, ALLC (always cooperate) and ALLD (always defect).  This cost involves a cognitive cost (to memorise previous interaction outcomes with co-players and make a decision based on them) and moreover, a cost of revealing the actual actions  of co-players (cf.~Introduction). The latter cost is usually ignored in previous works of IPD, but it can be non-trivial and thus significantly influence the nature of interactions. For instance, this cost is crucial to be considered in the context of human-machine interactions. For example, it might be quite costly and time consuming to check if one was charged the right amount when pay online/by Card/on ATM/ and whether the quality of the coffee produced by your coffee machine is reducing (and to what extent). This cost is even greater when interacting with intelligent agents whose operation and hence goals are less transparent, and which might, for example, be designed to hide pertinent information from users. 

\subsubsection*{Trust-based strategies}
\begin{figure}
    \centering
    \includegraphics[width=\linewidth]{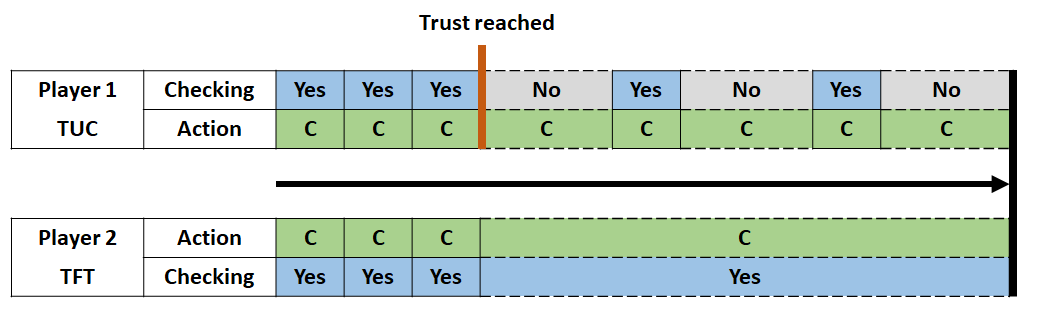}
    \caption{Diagram representing repeated interactions between a trust-based cooperator TUC and a tit-for-tat TFT. First, both strategies cooperate and check other player's action. After $\theta$ rounds (here $ \theta = 3$), trust is reached and TUC now checks the action of TFT occasionally with a probability $p$. Because TFT continues to cooperate, TUC continues to trust and to cooperate.}
    \label{fig:TUCvsTFT}
\end{figure}

We consider a new trust-based strategy that is capable of switching off the costly deliberation process when it trusts its co-players enough \footnote{Our modelling approach is in accordance with the definition of trust adopted in various multi-agent research, e.g.  \citep{dasgupta2000trust,ramchurn2004trust}. That is, trust is a belief an agent has that the other party will do what it says it will (being honest and reliable) or reciprocate (being reciprocal for the common good of both), given an opportunity to defect to get higher payoffs. }. Namely, this strategy starts an IPD interaction as a TFT player. When its ongoing trust level towards the co-player---defined here as the difference between the number of cooperative and defective moves from the co-player so far in the IPD---reaches a certain threshold, denoted by $\theta$, it will play C unconditionally. We denote this strategy by TUC. TUC is illustrated in the Figure \ref{fig:TUCvsTFT} representing one game between TUC and TFT.

\begin{figure}
    \centering
    \includegraphics[width=\linewidth]{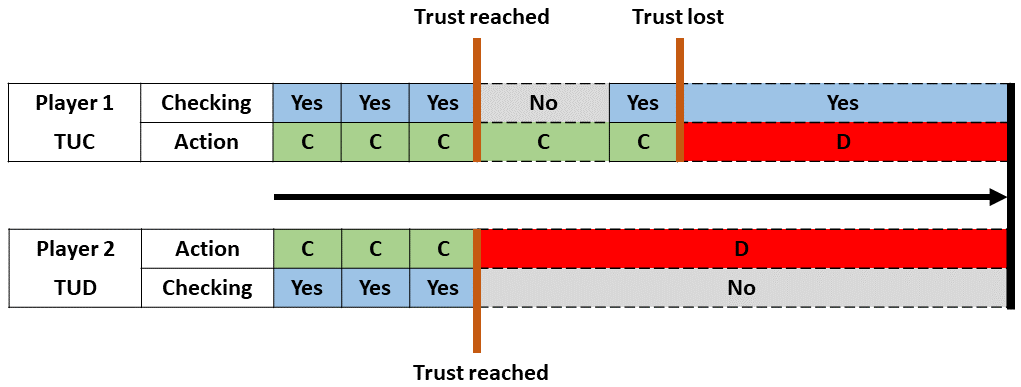}
    \caption{Diagram representing repeated interactions between a trust-based cooperator TUC and a trust-based defector TUD. First, both strategies cooperate and check the other player's action. After $\theta$ rounds (here $\theta = 3$), trust is reached for both strategies. TUC now cooperates and TUD defects. This continues until TUC checks and realises that TUD defects. After that, TUC looses trust, plays as a TFT and defects.}
    \label{fig:TUCvsTUD}
\end{figure}

Given the possibility of being exploited, but still to avoid costly deliberation, we assume that TUC will check, with a probability $p$, the co-player's actions after switching off \footnote{We assume that, given a sufficient cost of checking, TUC can always correctly find out the co-player's actions.}. If the co-player is found out to defect, TUC will revert to its initial strategy, i.e. TFT. 
As a counterpart of TUC, we consider TUD that whenever the ongoing trust level reaches the threshold $\theta$, switches to playing D unconditionally. TUD is illustrated in the Figure \ref{fig:TUCvsTUD} representing one game between TUC and TUD.

The payoff matrix for the five strategies ALLC, ALLD, TFT, TUC and TUD, can be given as follows 
{\small 
\begin{equation}
\label{Eq:payofmatrix}
\bordermatrix{	
	~ 		& \textbf{AllC} & \textbf{AllD} &	\textbf{TFT} & 
	\textbf{TUC}	& \textbf{TUD}	 \cr
	\textbf{AllC} 	& R %
					& S %
					& R %
					& R %
					& \frac{\theta R + (r - \theta)S}{r} %
					\cr
	\textbf{AllD} 	& T %
					& P %
					& \frac{T + (r-1)P}{r} %
					& \frac{T + (r-1)P}{r} %
					& \frac{T + (r-1)P}{r} %
					 \cr
	\textbf{TFT} 	& R - \varepsilon %
					& \frac{S + (r-1)P}{r} - \varepsilon %
					& R - \varepsilon %
					& R - \varepsilon %
					& \frac{\theta R + S + (r - \theta - 1) P}{r} - \varepsilon
					\cr %
	\textbf{TUC}	& R - \frac{\theta\varepsilon}{r} - \frac{p(r- \theta) \varepsilon}{r} %
					& \frac{S + (r-1)P}{r} - \varepsilon %
					& R - \frac{\theta \varepsilon}{r} - \frac{p(r- \theta) \varepsilon}{r} %
					& R - \frac{\theta \varepsilon}{r} - \frac{p(r- \theta) \varepsilon}{r} %
					& \Pi_{TUC,TUD} %
					\cr
	\textbf{TUD} 	& \frac{\theta R + (r - \theta)T - 
						\theta\varepsilon}{r} %
					& \frac{S + (r - 1)P}{r}-\varepsilon %
					& \frac{\theta R + T + (r - \theta-1)P - \theta 
						\varepsilon}{r} %
					& \Pi_{TUD,TUC}
					& \frac{\theta R + (r - \theta) P - 
						\theta\varepsilon}{r} %
					\cr
}
\end{equation}
}

For clarity, we write the payoff of TUC against TUD and TUD against TUC separately. Namely, the payoff of TUC against TUD is given by 
\begin{align*}
    & \Pi_{TUC,TUD} = \\ 
    &   \frac{1}{r}\left( \theta R - \theta \varepsilon + S + p (r^\prime-1) (P - \epsilon) + (1-p)(S + p(r^\prime-2) (P - \epsilon) + (1-p)[....]  ) \right) \\
    &= \frac{1}{r}\left(\theta R - \theta \varepsilon+ S \sum_{i = 0}^{r^\prime -1} (1-p)^i + p(P-\epsilon)\sum_{i = 0}^{r^\prime-1 } (r^\prime - i -1) (1-p)^i \right) \\
    &= \frac{\theta R - \theta \varepsilon}{r} + \frac{1}{r}\left(\frac{S(1-(1-p)^{r- \theta })}{p} + \frac{(P - \varepsilon)((1-p)^{r - \theta} + (r - \theta)p - 1))}{p}\right)
\end{align*}
where $r^\prime = r - \theta$. 
Similarly, 
\begin{align*}
    \Pi_{TUD,TUC} =\frac{\theta R - \theta \varepsilon}{r} +  \frac{1}{r}\left(\frac{T(1-(1-p)^{r - \theta }}{p} + \frac{P((1-p)^{r-\theta  } + (r- \theta )p - 1)}{p}\right)
\end{align*}
The payoff formulas can be explained as follows. In the first $\theta$ rounds both TUC and TUD play C and keep checking, so they obtain in each round $R - \epsilon$. As trust is reached, from next rounds TUC will check only occasionally with probability $p$. For example, if in the next round TUC checks, it obtains $S$ in that round and $P-\epsilon$ in the remaining rounds since it will play TFT. Otherwise, i.e. if TUC does not check in that round (with probability $1-p$),  the  process above is iterated for the payoffs calculation.   
\subsection{Evolutionary dynamics in finite populations}
We resort in this paper to Evolutionary Game Theory methods for finite populations understanding evolutionary dynamics of trust-based behaviours, in relation to other strategies \citep{key:imhof2005}. In this context, agents'  payoff represents their \emph{fitness} or social \emph{success}, and  evolutionary dynamics is shaped  by social learning \citep{key:Sigmund_selfishnes}, assuming that more successful agents will tend to be imitated more often by the others. 
We adapt here the pairwise comparison rule \citep{traulsen2006} to model social learning, where an agent $A$ with fitness $f_A$ adopts the strategy of another agent $B$ with fitness $f_B$ with probability given by the Fermi function, 
$$P_{A \rightarrow B} = \left(1 + e^{-\beta(f_B-f_A)}\right)^{-1}.$$ 
The parameter  $\beta$ stands for  the imitation strength or intensity of selection, i.e., how strongly agents base their decision to imitate on fitness comparison, where with $\beta=0$, the imitation decision is random, while for increasing $\beta$, imitation becomes increasingly deterministic.
 
In the absence of  behavioural exploration or mutations, end states of evolution inevitably are monomorphic. That is, whenever such a state is reached, it cannot be escaped via imitation. Thus, we further assume that, with some mutation probability, an agent can  freely explore its behavioural space.  In the limit of small mutation rates, the behavioural dynamics can be conveniently described by a Markov Chain, where each state represents a monomorphic population, whereas the transition probabilities are given by the fixation probability of a single mutant. The resulting Markov Chain has a stationary distribution, which characterises the average time the population spends in each of these monomorphic end states.

Suppose there exist at most two strategies in the population, say, $k$ agents using strategy A ($0 \leq k \leq N$)  and $(N-k)$ agents using strategies B. 
Let us denote by $\pi_{X,Y}$  the payoff an agent using strategy  $X$ obtained in an interaction with another individual using strategy  $Y$ (as given in the payoff matrix (\ref{Eq:payofmatrix})).
Hence, the (average) payoff of the agent that uses  A and B can be written as follows, respectively, 
\begin{equation} 
\label{eq:PayoffB}
\begin{split} 
\Pi_A(k) =\frac{(k-1)\pi_{A,A} + (N-k)\pi_{A,B}}{N-1},\\
\Pi_B(k) =\frac{k\pi_{B,A} + (N-k-1)\pi_{B,B}}{N-1},
\end{split}
\end{equation}

Now, the probability to change the number $k$ of agents using strategy A by $\pm$ one in each time step can be written as \citep{traulsen2006} 
\begin{equation} 
T^{\pm}(k) = \frac{N-k}{N} \frac{k}{N} \left[1 + e^{\mp\beta[\Pi_A(k) - \Pi_B(k)]}\right]^{-1}.
\end{equation}
The fixation probability of a single mutant with a strategy A in a population of $(N-1)$ agents using B is given by \citep{traulsen2006,Karlin:book:1975,key:imhof2005}
\begin{equation} 
\label{eq:fixprob} 
\rho_{B,A} = \left(1 + \sum_{i = 1}^{N-1} \prod_{j = 1}^i \frac{T^-(j)}{T^+(j)}\right)^{-1}.
\end{equation} 
When considering a set  $\{1,...,s\}$ of distinct strategies, these fixation probabilities determine the Markov Chain transition matrix $M = \{T_{ij}\}_{i,j = 1}^s$, with $T_{ij, j \neq i} = \rho_{ji}/(s-1)$ and  $T_{ii} = 1 - \sum^{s}_{j = 1, j \neq i} T_{ij}$. The normalised eigenvector of the transposed of $M$ associated with the eigenvalue 1 provides the above described  stationary distribution  \citep{key:imhof2005}, which defines the relative time the population spends while adopting each of the strategies.

\section{Results}

We use the  model defined above to answer two questions. First, when will individuals use trust? To answer this question, we investigate under which conditions trust is an evolutionary viable strategy. We measure the success of trust by the frequency of the trust-based cooperative strategy (TUC), i.e the proportion of time the population is composed of only TUC. Second, when should there be trust, in terms of how well the prevalence of trust-based behaviour enhances cooperation outcomes? We investigate the second question by looking under which conditions the presence of trust-based strategies (both TUC and TUD) increase the frequency of cooperation in the population. 

The default values of the parameters, unless otherwise specified,
are for the game payoffs $R = 1, S = -1, T = 2, P = 0$ (i.e. $b = 2$ and $c = 1$ in Donation game), the importance of the game $\gamma = 1$, the number of rounds $r = 50$, the population size $N = 100$, the intensity of selection  $\beta = 0.1$, the trust threshold $\theta = 3$, the probability of checking its partner $p = 0.25$ and the opportunity cost $\epsilon = 0.25$. The analysis of the model has been implemented using the package EGTTools \citep{Fernandez2020}.

\subsection{Trust as a mechanism to reduce opportunity costs}

\begin{figure}
\centering
\includegraphics[width=\linewidth]{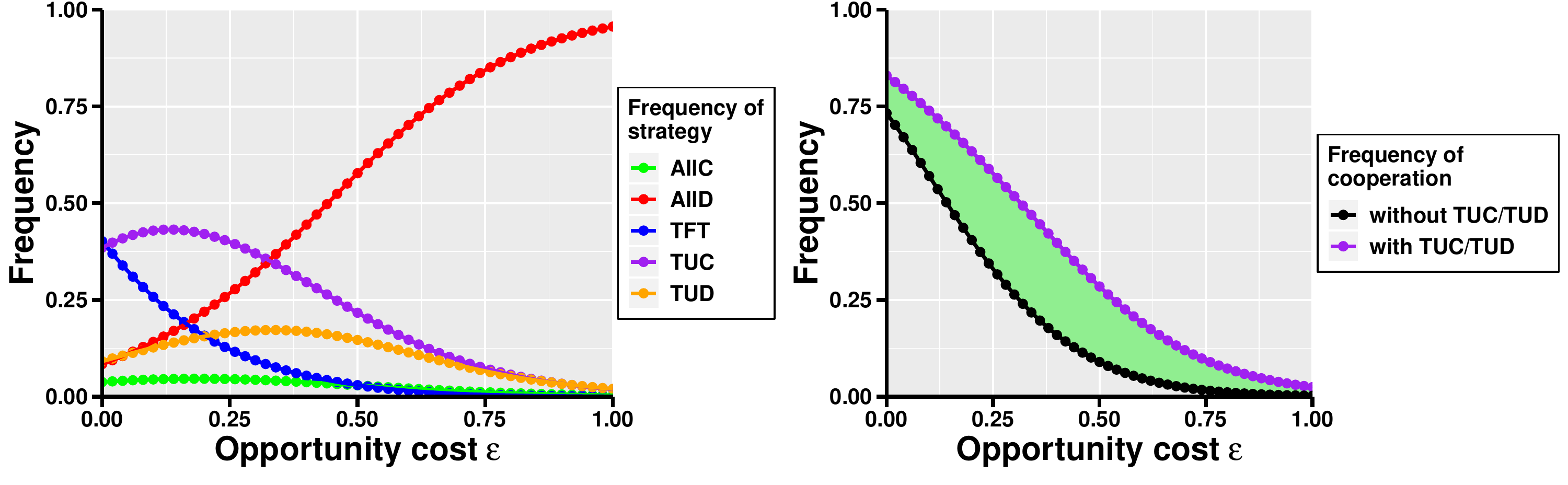}
\caption{\textbf{Left: Frequency of strategies as a function of the opportunity cost $\epsilon$. Right: Frequency of cooperation in absence or presence of trust-based strategies TUC and TUD, as a function of the opportunity cost $\epsilon$}.  The difference in frequency of cooperation between the two scenario is shaded in green when positive and red when negative.}
\label{fig:epsilon}
\end{figure}

The intuitive benefit of trusting something is to limit the cost of monitoring their actions in a long-term interaction, providing a shortcut in the decision making process. This is in line with several common definitions and theories of trust \citep{Luhmann:1979:a,Grabner-Kraeuter:2002:a,Petruzzi:2014:a}. Thus, we explore first the effect of opportunity cost $\epsilon$ on the strategies employed by individuals and the resulting frequency of cooperation.

The left panel of Figure \ref{fig:epsilon} shows that TUC is the most common strategy for a low to intermediate opportunity cost $\epsilon$ (between 0 and 0.3). When the opportunity cost $\epsilon$ is zero, both TUC and TFT are successful strategies and the population is composed of either one of them for most of the time. The success of TUC and TFT is explained by the capacity of these strategies to maintain high levels of cooperation within their homogeneous populations, while avoiding exploitation by AllD. Yet, the success of TFT is limited by the opportunity cost paid to check its partner's actions. This is shown in the results by the population being mostly AllD when the opportunity cost $\epsilon$ is high. Compared to TFT, TUC can limit this opportunity cost by reducing its attention to its partner's actions once trust is reached. This is why as the opportunity cost increases, the frequency of TFT plummets while TUC becomes more commonly observed. 
 
The right panel of Figure \ref{fig:epsilon} shows that the presence of trust-based strategies increases the frequency of cooperation. Importantly, this increase happens even when the opportunity cost $\epsilon$ is high ($\epsilon \approx c$), and not only when TUC is the most frequent, e.g. for low $\epsilon$. This is because a high frequency of cooperation is already reached for a low opportunity cost due to TFT. The presence of TUC has a more important effect on cooperation when the opportunity cost increases since in that case the performance of TFT significantly reduces.

To conclude, trust-based cooperation is a particularly common strategy, in particular in interactions with moderate opportunity cost, and it promotes cooperation for a large range of opportunity costs.

\subsection{Length of interactions and importance of the game}

\begin{figure}
\centering
\includegraphics[width=\linewidth]{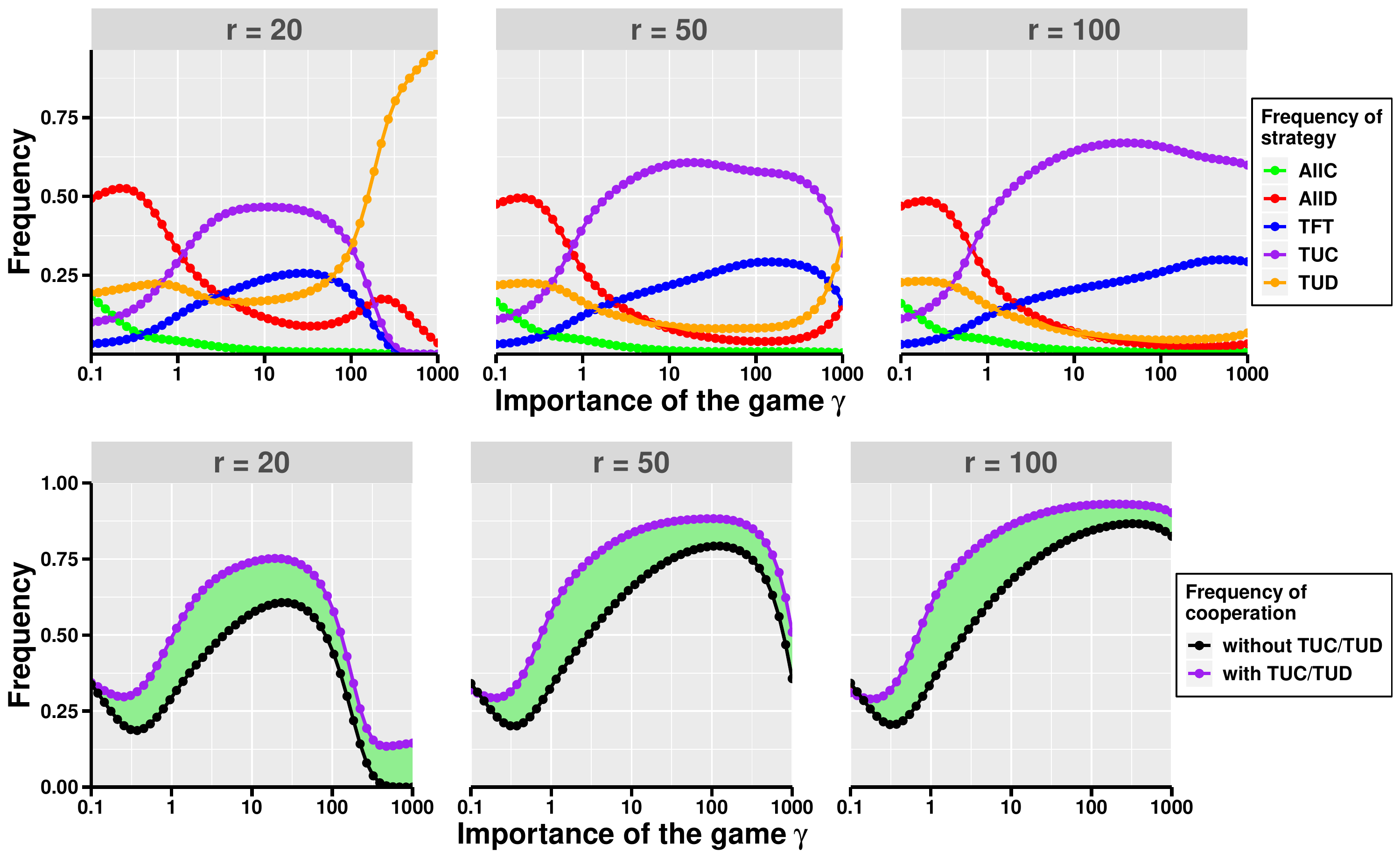}
\caption{\textbf{Top: Frequency of strategies as a function of the number of rounds $r$ and importance of the game $\gamma$ (logarithmic scale);  Bottom: Frequency of cooperation in absence or presence of trust-based strategies TUC and TUD, as a function of the number of rounds $r$ and importance of the game $\gamma$ (logarithmic scale).} For clarity, the difference in frequency of cooperation is shaded in green when positive and red when negative.}
\label{fig:r_gamma}
\end{figure}

We now investigate (i) the importance of the game $\gamma$ because this affects the \textit{relative} cost of checking the other player and (ii) the number of rounds, because this affects the \textit{relative} time that is required for trust to be established. The results are presented in Figure \ref{fig:r_gamma}. First, we discuss the cases on the left column, where repeated interactions are short (expected number of rounds $r = 20$). The top left panel of Figure \ref{fig:r_gamma} shows that in such conditions, TUC is successful for medium importance of the game. TUC is also the most frequent strategy for a large range of importance of the game (note that the results presented are on a logarithmic scale). When the importance of the game is very low e.g. $\gamma = 0.1$, the most frequent strategy is AllD. In this condition, the opportunity cost is too high relative to the benefit provided by cooperation for either of the conditionally cooperative strategies, TUC or TFT, to thrive. When the importance of the game is very high, e.g. $\gamma = 1000$, TUC is almost never observed and TUD is, by far, the most frequent strategy. When the importance of game is high, defecting while the other player cooperates provides a huge benefit. AllD gets this benefit on the first round played with AllC,  TFT and TUC. On the other hand, TUD  obtains this advantageous payoff at least on the round after trust is established  when interacting with TUC and TFT. This advantage by TUD is hard to recover through reciprocity if the number of rounds is not sufficiently high.

This result is dependent of the length of the interactions. The top part of Figure \ref{fig:r_gamma} shows that a higher number of rounds $r$ leads to (i) a higher frequency of TUC and (ii) the prevalence of TUC for a wider range of importance of game $\gamma$. TUC remains the most frequent strategy even when the importance of the game is high if the interactions are sufficiently long. This is because the high number of rounds where both individuals cooperate make up for the few initial rounds where TUC is exploited by TUD (and on a lesser extent, AllD).

The bottom part of Figure \ref{fig:r_gamma} shows that the presence of trust-based strategies increases the frequency of cooperation for all conditions examined. The highest frequency of cooperation is obtained for long interactions and high importance of the game. As shown by the similar shape of the curves, the higher frequency of cooperation appears to result from the high frequency of TUC. There is one notable exception. As shown in the bottom left figure (low $r$ and high $\gamma$), the presence of trust based strategies also increases cooperation when TUC is not present. This is because TUD strategies cooperate more (for $\theta$ rounds) than AllD strategies which never cooperate.

In conclusion, trust is favoured for long-term interactions and can be observed on a wide range of importance of the game. The presence of trust-based strategies increases the frequency of cooperation for the whole set of parameter values studied.

\subsection{Trustfulness and TUD}

\begin{figure}
\centering
\includegraphics[width=\linewidth]{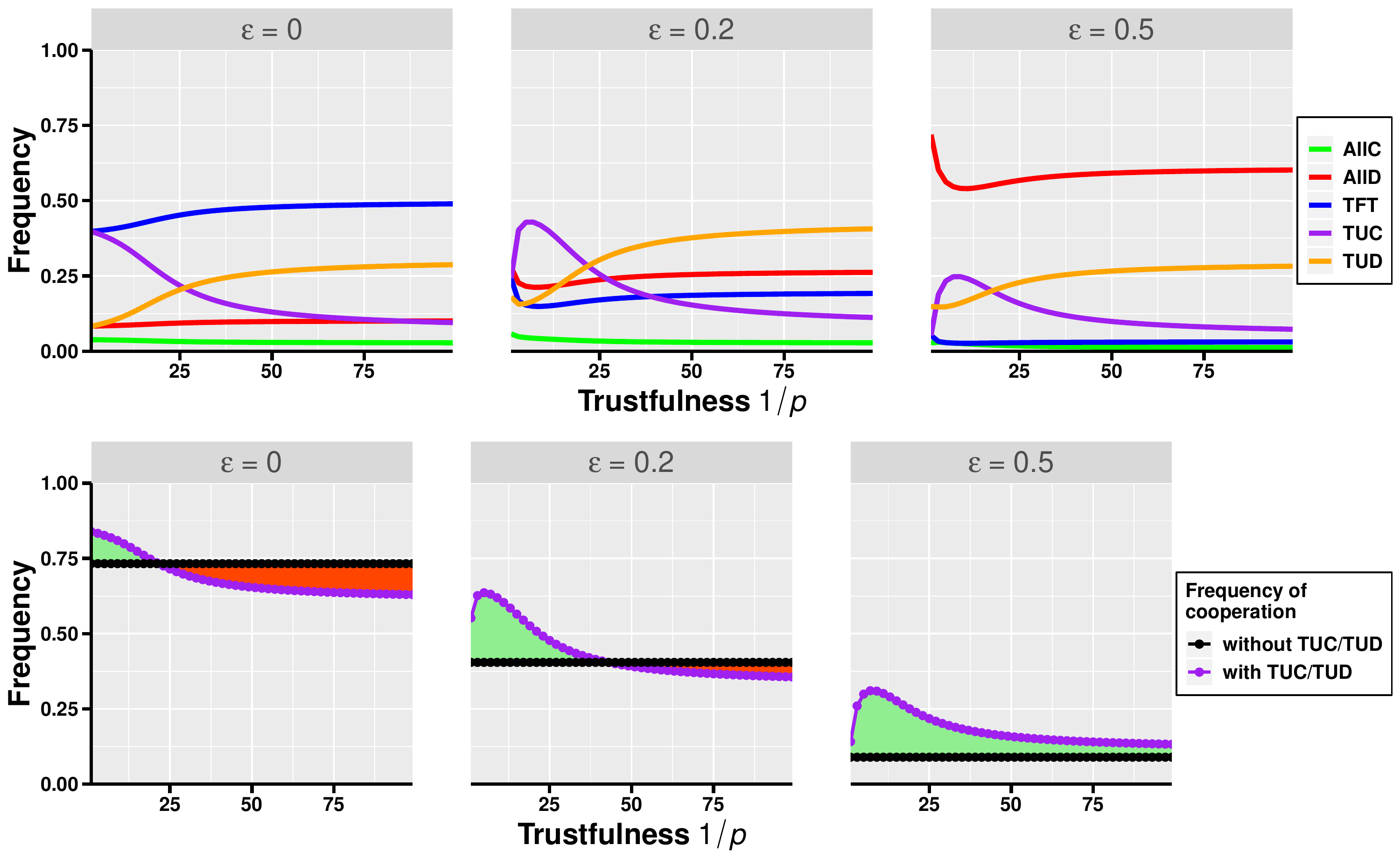}
\caption{\textbf{Top: Frequency of strategies as a function of the opportunity cost $\epsilon$ and trustfulness $1/p$ (average number of rounds between checking event). Bottom: Frequency of cooperation in absence or presence of trust-based strategies TUC and TUD, as a function of the opportunity cost $\epsilon$ and trustfulness $1/p$ (average number of rounds between checking event)}. For clarity, the difference in frequency of cooperation is shaded in green when positive and red when negative.
}
\label{fig:p_epsilon}
\end{figure}

We have seen from the above results that trust-based cooperators are vulnerable to exploitation by TUD players, which are specifically tailored to take advantage of unconditional trust. This vulnerability was limited so far as we considered a rather careful truster with a $p = 0.25$. We now look at what is the effect of the probability of checking $p$ on the success of TUC and the frequency of cooperation. For clarity, we present the result as a function of $1/p$, which approximates the trustfulness  (which is larger for a smaller probability of checking) of TUC on the overall game, rather than $p$, which represents the carefulness of TUC on a single round.

The top part of Figure \ref{fig:p_epsilon} first confirms that it is important for TUC's success to continue checking after trust is reached as TUC is much less frequent for a high value of trustfulness (i.e high $1/p$). If TUC is too trustful, the game is either dominated by TFT when the opportunity cost is small, by TUD when the opportunity cost is intermediate, and and AllD when the opportunity cost is high. There is an intermediate optimal trustfulness $1/p$ at which TUC is the most frequent strategy (except for zero opportunity costs where the lowest trustfulness and the highest probability of checking is the best strategy, which is equivalent to TFT). On the one hand, low trustfulness makes TUC less successful because TUC checks its partner often and so pays a higher opportunity cost. On the other hand, high trustfulness makes TUC vulnerable to exploitation by TUD for a longer number of rounds. The results show that there can be an optimal level of trust resulting from this trade-off.

The bottom part of Figure \ref{fig:p_epsilon} shows that the presence of trust-based strategies increases the frequency of cooperation when the opportunity cost is moderate or high. This cooperation improvement is the highest for the optimal trustfulness at which TUC is very frequent. Again, the presence of trust-based strategies can lead to an increase in the frequency of cooperation even if they are not the most frequent strategy e.g. for very high opportunity costs $\epsilon$.
Unlike previously, the results also show that the presence of trust-based strategies can reduce the frequency of cooperation. This happens when the opportunity cost $\epsilon$ is low and the trustfulness $1/p$ is high. In these conditions, trustful and careless TUC players get exploited by TUD players, which increases the frequency of TUD, making cooperation a less viable option (evolutionarily). In the absence of trust-based strategies, TFT is careful enough to avoid this pitfall. 

To conclude, unconditional trust is a viable strategy only in limited conditions and how much TUC relies on trust can have significant effect on the success of the strategy.

\section{Discussion}

Trust is a commonly observed mechanism in human interactions, and discussions on the role of trust are being extended to social interactions between humans and intelligent machines \citep{Andras2018TrustingSystems}. It is therefore important to understand how people behave when interacting with those machines; particularly, whether and when they might exhibit trust behaviour towards them? Answering these questions is crucial for providing suitable designs of mechanisms or infrastructures to facilitate human-intelligent machine interactions, e.g. in engineering pro-sociality in a hybrid society of humans and machines \citep{paiva2018engineering}. To this end, this paper provides a game theoretic analysis, where we formalised trust as a behavioural strategy and integrated it into an EGT model to study (i) its success in competition with non-trusting strategies and (ii) its effect on the level of cooperation.

Our results show first that trust is expected to be a pervasive cooperation enabling strategy. It is a frequent strategy for a large range of parameters, even in the presence of other strategies that are traditionally successful such as unconditional defection and Tit-for-Tat. Second, our results show that trust is a desirable mechanism in social systems because the presence of trust-based strategies increases the level of cooperation for a wide range of parameters. Finally, we show that trust-based cooperators are vulnerable to trust-based defectors, which are specialised to exploit them. However, our results also suggest that a minimum carefulness after trust is reached (low $p$) strongly limits this vulnerability. 

Overall, our analysis  shows that trust can emerge because it reduces the opportunity costs paid by individuals during interactions. It is a form of cognitive shortcut that, while exposing the player to some risks, can allow individuals to cooperate at lesser cost. If the pitfalls of trust have often been discussed, our results underlie the importance of taking into account both the benefits and the risk that the use of trust involves. Understanding the balance between these two is a first step to optimise the benefits of trust in intelligent machines while limiting the costs. On this line, further work could expand the model to look at different forms of trust based cooperation and defection strategies, how they co-evolve, and how exploitation of trust-based cooperators can be avoided. It is noteworthy that our additional (numerical) analyses have shown  that all the observations described  above (i.e. in Figures \ref{fig:epsilon} -  \ref{fig:p_epsilon}) are robust, e.g. for different  values of the threshold number of rounds required for trust to be established (i.e. $\theta$) (see Appendix). Moreover, note that in the current model we consider that TUC and TUD have the same $\theta$, which is the worst case scenario for the evolution of TUC (and cooperation), since it represents the situation where TUD can perfectly recognise when TUC starts trusting a cooperative co-player and therefore becomes less vigilant of exploitation. More realistically, TUD might need to spend extra resource to gather information about TUC (e.g. providers learn about their customers' preferences and behaviours) to determine what is TUC's $\theta$. On the other hand, TUC should not easily reveal or make available their information (that can be used to infer their $\theta$), to better deal with TUD. Future work should address how these aspects might change the outcome of the evolutionary dynamics.

One of the most famous previous formalisations of trust is an experiment from behavioural economics called the trust game \citep{Berg1995TrustHistory}. This game consists of one individual receiving an endowment of money, of which it must choose a certain amount (which can be zero) to send to the other player. The amount sent to the other player is tripled by the experimenter (so that sending money represents an investment). The other player then decides what amount of this money (if any) to send back to the first player (so that there is risk in the first player sending money). While the Nash equilibrium is for the first player to send nothing to the other player, in experiments individuals usually deviate from this by sending a positive amount \citep{Berg1995TrustHistory}. The amount that the first player sends can be understood as measuring how much the first player trusts the second to reciprocate. 
However, in contrast to our formalisation, the trust game measures more a willingness to take risks blindly, as interactions are between anonymous individuals and are played only once. By contrast, we have considered repeated interactions between the same individuals, which has enabled us to  look at the success of strategies that build up trust over time. 

In line with our approach, trust enabling strategies were previously considered in the context of repeated games \citep{han2011intention}, where trust is built over time as a component of a larger decision making process, for prediction of opponents' behaviour. Trust was also considered for enabling cooperation in a one-shot prisoner's dilemma  \citep{janssen2008evolution,mcnamara2009evolution}, but it was assumed that players can recognise how trustworthy a co-player is based on additional cues such as signalling. 
Our work differs from these approaches in that we consider trust as a cognitive shortcut to avoid deliberation and having to check the outcomes of previous interaction(s), thereby limiting the opportunity cost of conditional strategies. 

In addition, trust has been used extensively in various computerised systems, such as in multi-agent open and distributed systems, for facilitating agents' interactions. Agents may have limited computational and storage capabilities that restrict their control over interactions, and trust is used to minimise uncertainty associated with the interactions, especially when agents inhabit in uncertain and constantly changing environments \citep{ramchurn2004trust,falcone2001social}. This is the case for various applications including peer-to-peer computing, smart-grids, e-commerce, etc \citep{kamhoua2011game,ramchurn2004trust,papadopoulou2001trust,Petruzzi:2014:a,Brooks:2020:a}. These studies utilise trust for the purpose of regulating individual and collective behaviours, formalising different aspects of trust (such as reputation and belief) \citep{Castelfranchi1997ModelingAgents,castelfranchi2010trust}. 
Our results and approach provide novel insights into the design of such computerised and hybrid systems as these require trust to ensure high levels of cooperation or efficient collaboration within a group or team of agents, including human-machine hybrid interactions. For instance, our results show that the importance of the business at hand (relative to the opportunity cost) needs to be taken into account to ensure a desired level of cooperation. Also, the system needs to be designed so that the opportunity cost of verifying the actions of an intelligent machine is sufficiently low (relative to the benefit and cost of the game) to enable a long-term trusting relationship with customers, e.g. making the activities transparent either directly to the user or to expert auditors that follow professional codes of ethics \citep{Andras2018TrustingSystems}.

In the current work we have focused on the Prisoner's Dilemma as it represents the hardest (pairwise) scenario for cooperation to emerge. Many other scenarios, such as the intelligent virtual assistant planning the route to work example in the Introduction, might be represented using other social dilemmas such as coordination or snowdrift games \citep{key:santos2006,key:Sigmund_selfishnes}. 
Considering such games where it is easier for cooperation to emerge has the potential to open new windows of opportunity for long-term trust-based relationships to be established.
Our future work will study how trust-based strategies (as we have modelled) evolve in the context of other social dilemmas.

\section{Conclusion}

We have demonstrated in this paper that evolutionary game theory provides a valuable framework to study trust. Social interactions often result in complex dynamics with unexpected consequences, which a quantitative model is able to shed light on. Our model provides formal support for the theory that trust is a cognitive shortcut which people use to reduce the complexity of their interactions. The results of the model provide new insights into the questions of whether and when humans might trust intelligent machines, generating reasonable behavioural hypotheses that empirical studies can test. 

\section*{Acknowledgements}
We thank all members of the TIM 2019 Workshop, particularly, Lukas Esterle and Stephen Marsh, for useful discussion. TAH and CP are supported by Future of Life Institute (grant RFP2-154).

\newpage
\section*{Competing interests statement} \noindent The authors have not competing interests. \\

\section*{Author contributions}
\noindent \textbf{The Anh Han}: Conceptualization, Methodology, Formal Analysis, Writing - Original draft preparation \\
\textbf{Cedric Perret}: Conceptualization, Methodology, Formal Analysis, Software, Writing - Original draft preparation \\
\textbf{Simon T. Powers}: Conceptualization, Methodology, Writing - Original draft preparation \\

\clearpage 
\section*{Appendices}
\subsection*{Appendix A: Results for different trust threshold $\theta$}

\begin{figure}
\centering
\includegraphics[width=\linewidth]{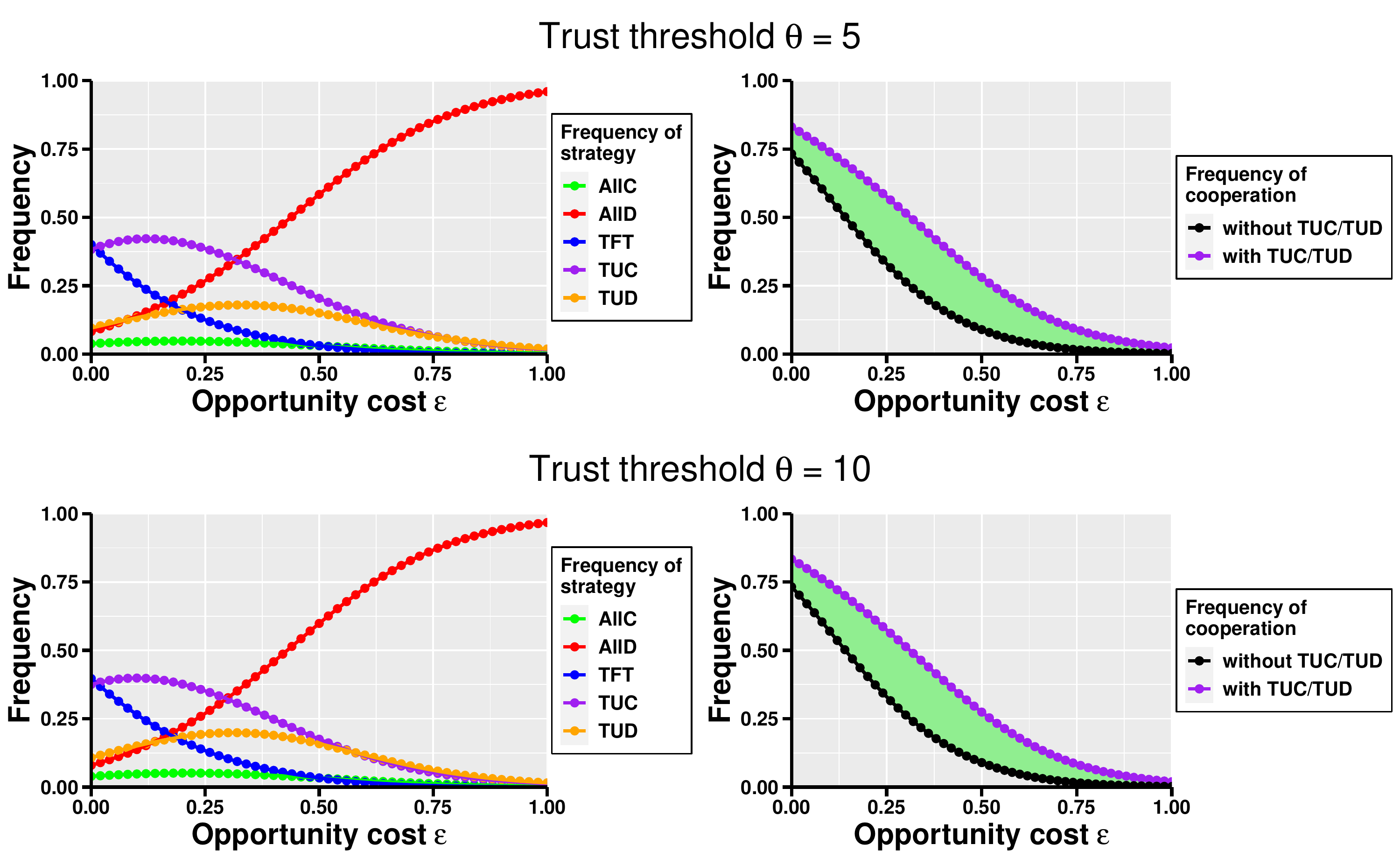}
\caption{\textbf{Left: Frequency of strategies as a function of the opportunity cost $\epsilon$. Right: Frequency of cooperation in absence or presence of trust-based strategies TUC and TUD, as a function of the opportunity cost $\epsilon$}.  The difference in frequency of cooperation between the two scenario is shaded in green when positive and red when negative. Each results are presented for different trust threshold $\theta = 5$ and $\theta = 10$. Parameters: $\beta = 0.1$, population size $N = 100$, $\gamma = 1$, $r = 50$, $p=0.25$,  $R = 1, S = -1, T = 2, P = 0$.}
\label{fig:epsilon_theta}
\end{figure}

\begin{figure}
\centering
\includegraphics[width=\linewidth]{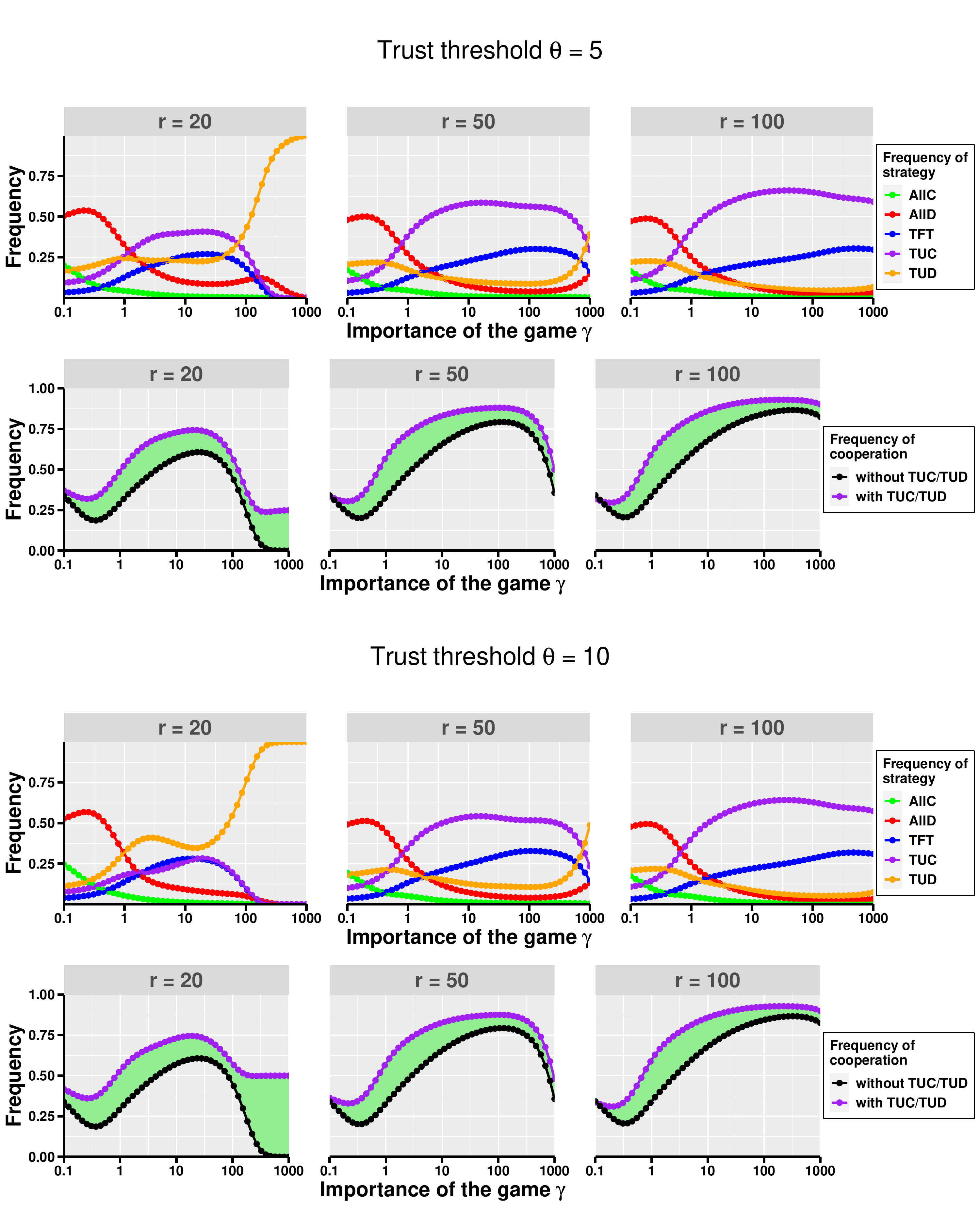}
\caption{\textbf{Top: Frequency of strategies as a function of the number of rounds $r$ and importance of the game $\gamma$ (logarithmic scale);  Bottom: Frequency of cooperation in absence or presence of trust-based strategies TUC and TUD, as a function of the number of rounds $r$ and importance of the game $\gamma$ (logarithmic scale).} For clarity, the difference in frequency of cooperation is shaded in green when positive and red when negative. Each results are presented for different trust threshold $\theta = 5$ and $\theta = 10$. Parameters: $\beta = 0.1$, population size $N = 100$, $p=0.25$,  $R = 1, S = -1, T = 2, P = 0$.}
\label{fig:r_gamma_theta}
\end{figure}

\begin{figure}
\centering
\includegraphics[width=\linewidth]{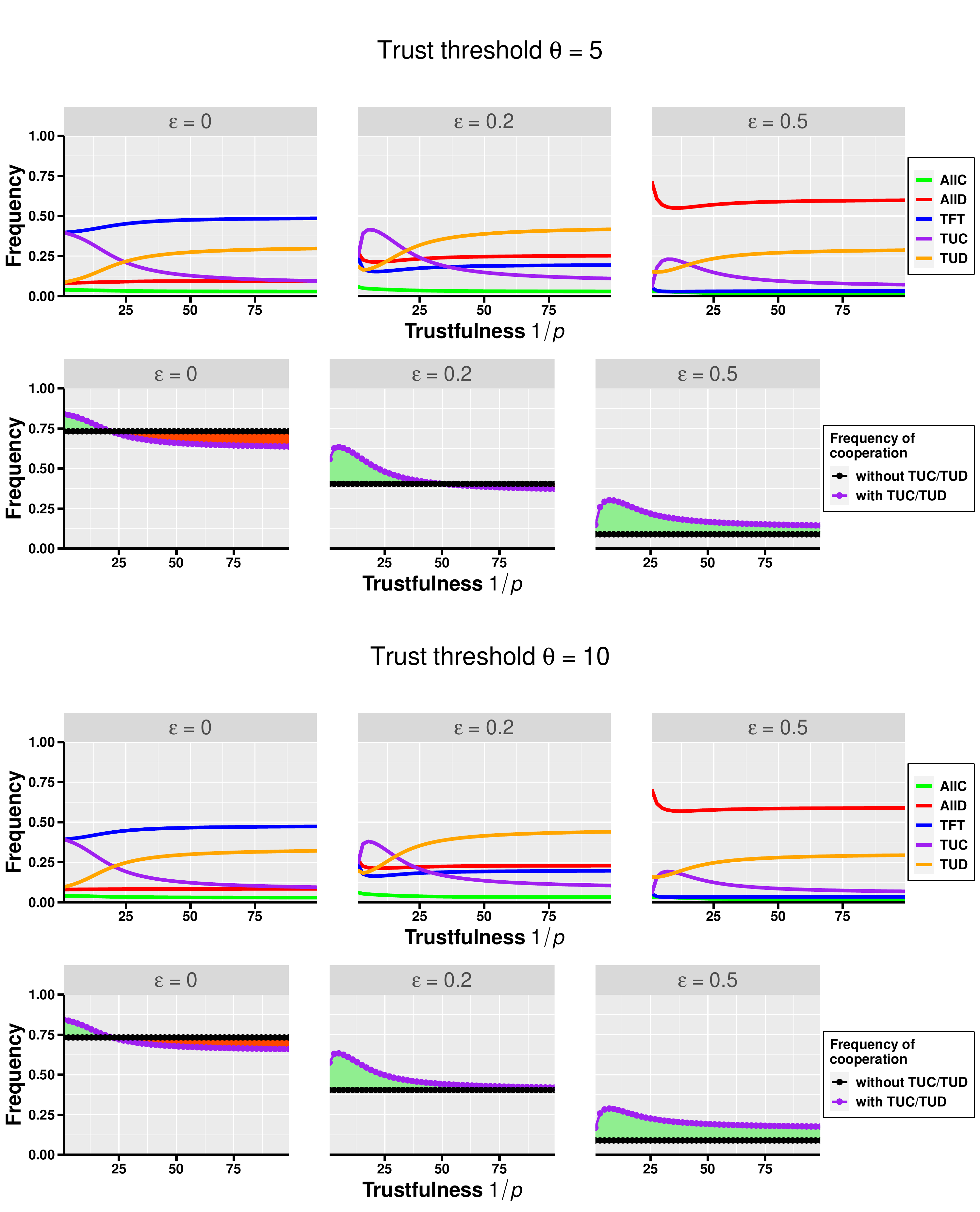}
\caption{\textbf{Top: Frequency of strategies as a function of the opportunity cost $\epsilon$ and trustfulness $1/p$ (average number of rounds between checking event). Bottom: Frequency of cooperation in absence or presence of trust-based strategies TUC and TUD, as a function of the opportunity cost $\epsilon$ and trustfulness $1/p$ (average number of rounds between checking event)}.  For clarity, the difference in frequency of cooperation is shaded in green when positive and red when negative. Each results are presented for different trust threshold $\theta = 5$ and $\theta = 10$. Parameters: $\beta = 0.1$, population size $N = 100$, $\gamma=1$,$r=50$, $R = 1, S = -1, T = 2, P = 0$.}
\label{fig:p_epsilon_theta}
\end{figure}

\end{document}